\documentclass[10pt,letterpaper,twocolumn]{article} %% two column, final layout

\usepackage{ol2}
               
\begin{document}

\twocolumn[ %% activate for two-column option

\title{Timing attacks on practical quantum cryptographic systems}

\author{Nitin Jain}

\address{Max Planck Institute for the Science of Light, G\"unther-Scharowsky-Str. 1/Bau 24, 91058 Erlangen, Germany \\
%Institut f\"ur Optik, Information und Photonik,
Universit\"at Erlangen-N\"urnberg, Staudtstra\ss e 7/B2, 91058, Erlangen, Germany \\
$^*$Email: nitin.jain@mpl.mpg.de 
}

\begin{abstract}
With photons being the only available candidates for long-distance quantum communication, most quantum cryptographic devices are physically realized as optical systems that operate a security protocol based on the laws of quantum mechanics. But to finally yield a stream of bits (secret key) usable for encryption, a quantum-to-classical transition is required. Synchronization of electronic \& optoelectronic components involved in such tasks thus becomes a necessary and important step. However, it also opens up the possibility of timing-based loopholes and attacks. 
\end{abstract}

 ] %% activate for two-column option

\section{Introduction} %
In a letter~\cite{Born69} to Max Born written in 1926, Albert Einstein remarked: ``Quantum mechanics is certainly imposing. But an inner voice tells me that this is not yet the real thing. The theory says a lot, but does not bring us any closer to the secrets of the Old One. I, at any rate, am convinced that He is not playing dice.'' This quote, particularly the last part about God not playing dice, indicates Einstein's unwillingness to accept a fundamental tenet of quantum theory: with regards to values of physical quantities, only statistical assertions can be permitted. Indeed, Einstein and some other prominent scientists were also inclined towards the more classical view of the world in which physical systems could be ascribed properties that existed irrespective of whether they were being measured or not~\cite{epr35}. It was thus believed by some that quantum mechanics could not provide a complete description of Nature. 

Fast forward to the next century, and with principles of quantum mechanics having been verified in innumerable different experiments, it seems that the earlier view of those scientists was incorrect. Nonetheless, due to its bizarre nature and ideas, quantum mechanics still confounds anyone who tries to understand it. But thankfully, that hasn't stopped us from exploring and devising applications that utilize non-classical world features, such as quantum entanglement and superposition. While elements of `quantum physics in action' have already been available to us on daily basis (e.g. barcode scanners, mobile phones), applications such as quantum computers, quantum cryptographic systems etc. that have much wider implications are now also getting deployed~\cite{swisselec07,dwave12}. The focus of this article is on quantum key distribution (QKD) -- the most successful and well-known application of quantum cryptography, as of today. So much so that they are frequently used interchangeably, and this shall also be the case with this article. 
\begin{figure}
\centerline{\includegraphics[width=8cm]{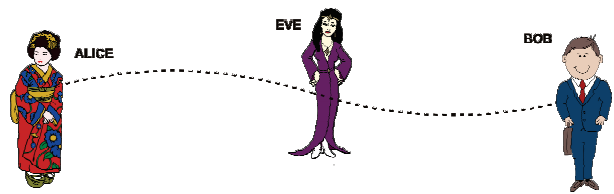}}
\caption{\textit{Fundamental scenario for cryptography}: Two entities, normally called Alice and Bob wish to share a secret, which a third party usually called Eve (often also harbouring a malicious intent) is interested in knowing.}
\label{figAliceBobEve}
\end{figure}

Figure~\ref{figAliceBobEve} illustrates the basic scheme of any cryptographic system. Two users Alice and Bob communicate privately (typically using a conveniently-available medium such as the Internet or telephone network) and an adversary Eve is interested in knowing the secret that they wish to share. Alice and Bob generally use a stream of bits called \emph{keys} to protect their communication from being eavesdropped. While conventional schemes use the computational hardness offered by public-key cryptography (PKC) to prevent such eavesdropping, quantum cryptography relies on the disturbances -- introduced by an eavesdropper -- in the communication of the secret key between Alice and Bob. 

These disturbances can be detected and quantified and indicate the ultimate level of security that can be obtained - this depends on the quantum cryptographic protocol employed. If the protocol runs successfully, Alice and Bob could use methods such as the one-time pad (OTP) for encrypting and decrypting the actual message (plain text) that they wish to share. A well-known property of OTP is its \emph{information-theoretical immunity to cryptanalysis}. Also, to successfully eavesdrop in this scenario, Eve must crack the secret key in real time. This is, however, not the case with PKC: previously encrypted messages (that may still be of vital importance, e.g. military plans) could be cracked if sufficient computational resources are at disposal and/or faster mathematical algorithms are discovered. With advances in quantum computing, the day is probably not far when this would actually happen! 
\section{A quantum-cryptographic protocol explained\ldots \, without any quantum mechanics}\label{sec:mbp}
I shall now endeavour to explain perhaps the most famous and widely-used QKD protocol. And without resorting to \emph{any} quantum mechanics! 
That sounds contradictory of course, because if it were possible to \emph{classically} explain the innards of a quantum cryptographic protocol, then why do physicists delve into the usually unusual quantum theories at all? So I have to make some unrealistic assumptions that will be good enough for capturing the essence of how the protocol works, but (obviously) do not suffice to cover all the quantum-mechanical aspects. Assume that there exists an \emph{unlimited supply of tiny magical balls} with the following characteristics:
\flushleft
\begin{enumerate}
\item[P1.] The state of a ball is encoded by a number $\mathbf{n}$; for operation of the protocol, it is chosen to be amongst the first four positive integers, i.e. $\mathbf{n} = \left\{\mathbf{1}, \mathbf{2}, \mathbf{3}, \mathbf{4}\right\}$. Note that exactly two out of the four numbers in this set are:
\begin{itemize}
\item Odd ($\mathbf{n} = \mathbf{1}$ and $\mathbf{3}$)
\item Even ($\mathbf{n} = \mathbf{2}$ and $\mathbf{4}$)
\item More than $\mathbf{2}$ ($\mathbf{n} = \mathbf{3}$ and $\mathbf{4}$)
\item Less than $\mathbf{3}$ ($\mathbf{n} = \mathbf{1}$ and $\mathbf{2}$)
\end{itemize}
We denote these four possibilities by keywords $K = \left\{OD, EV, M2, L3\right\}$, respectively. The encoding procedure does not take a number as the input, instead, it takes a keyword which is chosen in a random fashion from $K$. 
\item[P2.] Uttering a keyword to a ball makes it randomly pick a suitable value of $\mathbf{n}$ that satisfies the constraint imposed by the keyword. E.g., if the uttered keyword is $M2$, both $\mathbf{n}=\mathbf{3}$ and $\mathbf{n}=\mathbf{4}$ are equally-likely candidates. The ball then encodes itself with the picked value. Table \ref{tabMBPenc} shows an example of 12 different balls encoded one after the other. 
\item[P3.] Once encoded, a ball retains no memory of the keyword that was used to encode it. It does know $\mathbf{n}$ itself, but cannot be forced to reveal the value (even the person who encoded the balls cannot do it). However, the ball is programmed to truthfully and logically answer ``Yes'' or ``No'' (bit \textsl{1} and \textsl{0}, respectively) to one and only one of the following two questions. %``1'' and ``0''
\begin{itemize}
\item[\textit{a}.] Is $\mathbf{n}$ odd (denoted by $nOD$)? 
\item[\textit{b}.] Is $\mathbf{n}$ less than $\mathbf{3}$ (denoted by $nL3$)? 
\end{itemize}
Note, if a ball answers ``Yes'' to the above questions, it would correspondingly answer ``No'' to 
\begin{itemize}
\item[\textit{c}.] Is $\mathbf{n}$ even? 
\item[\textit{d}.] Is $\mathbf{n}$ more than $\mathbf{2}$? 
\end{itemize}
In other words, the possibility of $\mathbf{n}$ being more than $\mathbf{2}$ completely excludes that of $\mathbf{n}$ being less than $\mathbf{3}$ and vice versa. Similarly, the possibility of oddness complements the possibility of evenness. So it suffices to say that asking \textit{a} $\left[\rm \textit{b}\right]$ is equivalent to \textit{c} $\left[\rm \textit{d}\right]$. 
\item[P4.] An attempt to pry open the ball triggers a self-demolition procedure and the ball vanishes in a puff of smoke! Also, after answering a question the ball falls silent: no amount of prodding or poking, pleading or praying can make it talk.
\end{enumerate}
\begin{center}
\begin{table*} %TABLE FOR ENCODING OF MAGIC BALLS
%\centering
{\small
\hfill{}
%\centerline{\includegraphics[width=13cm]{
\begin{tabular}{|c|c|c|c|c|c|c|c|c|c|c|c|c|}
\hline
 & 1 &	2 &	3 &	4	& 5	& 6 &	7	& 8	& 9 &	10 & 11 & 12 \\
\hline
Keyword used & $OD$ &	$M2$ &	$L3$ & $EV$ &	$L3$ & $M2$ & $EV$ & $EV$ & $L3$ & $EV$ & $OD$ & $M2$ \\
\hline
Value of $\mathbf{n}$ & $\mathbf{3}$ &	$\mathbf{4}$ & $\mathbf{2}$ &	$\mathbf{2}$ & $\mathbf{1}$	& $\mathbf{3}$ & $\mathbf{4}$	& $\mathbf{2}$ & $\mathbf{1}$ &	$\mathbf{4}$ & $\mathbf{1}$ & $\mathbf{3}$ \\
\hline
\end{tabular}}
\hfill{} 
\caption{\textit{Encoding of the magical balls}. Balls randomly pick suitable values to encode themselves, e.g. the balls choose $\mathbf{4}$ and $\mathbf{2}$ when $EV$ is invoked in slots 7 and 8, respectively. However, in slots 9 and 11, invoking two different keywords $L3$ and $OD$ still results in the same value $\mathbf{n=1}$ encoded.}
\label{tabMBPenc}
\end{table*}
%\end{center} 
%\begin{center}
\begin{table*} %TABLE FOR MAGIC BALL PROTOCOL
%\centering
{\small
\hfill{}
%\centerline{\includegraphics[width=13cm]{
\begin{tabular}{|c|c|c|c|c|c|c|c|c|c|c|c|c|}
\hline
 & 1 &	2 &	3 &	4	& 5	& 6 &	7	& 8	& 9 &	10 & 11 & 12 \\
\hline
Keyword used & $\textcolor{green}{OD}$ &	$M2$ &	$\textcolor{green}{L3}$ & $\textcolor{green}{EV}$ &	$\textcolor{green}{L3}$ & $\textcolor{green}{M2}$ & $EV$ & $\textcolor{green}{EV}$ & $L3$ & $EV$ & $\textcolor{green}{OD}$ & $\textcolor{green}{M2}$ \\
\hline
Value of $\mathbf{n}$ & $\mathbf{3}$ &	$\mathbf{4}$ & $\mathbf{2}$ &	$\mathbf{2}$ & $\mathbf{1}$	& $\mathbf{3}$ & $\mathbf{4}$	& $\mathbf{2}$ & $\mathbf{1}$ &	$\mathbf{4}$ & $\mathbf{1}$ & $\mathbf{3}$ \\
\hline
Bob's question & $nOD$ & $nOD$ & $nL3$ & $nOD$ & $nL3$ & $nL3$ & $nL3$ & $nOD$ & $nOD$ & $nL3$ & $nOD$ & $nL3$ \\
\hline
Ball's answer & Yes &	No & Yes & No	& Yes	& No & No	& No & Yes & No & Yes & No \\
\hline
Bits (Bob) & \textsl{\textcolor{green}{1}} &	\textsl{0} & \textsl{\textcolor{green}{1}} & \textsl{\textcolor{green}{0}} & \textsl{\textcolor{green}{1}} & \textsl{\textcolor{green}{0}} & \textsl{0}	& \textsl{\textcolor{green}{0}} & \textsl{1} & \textsl{0} & \textsl{\textcolor{green}{1}} & \textsl{\textcolor{green}{0}}\\
\hline
\end{tabular}}
\hfill{} 
\caption{\textit{Magical ball protocol in action}. While Alice randomly encodes the balls, Bob randomly questions them. After discussing on a public channel, they are able to process their data such that they finally obtain the same secret key (bit sequence in green) with a high probability.}
\label{tabMBP}
\end{table*}
\begin{table*} %TABLE FOR FOILING OF EVE
{\small
\hfill{}
%\centerline{\includegraphics[width=13cm]{
\begin{tabular}{|c|c|c|c|c|c|c|c|c|c|c|c|c|}
\hline
 & 1 &	2 &	3 &	4	& 5	& 6 &	7	& 8	& 9 &	10 & 11 & 12 \\
% & \textcolor{green}{1} &	2 &	\textcolor{green}{3} & \textcolor{green}{4}	& \textcolor{red}{5} & \textcolor{green}{6} & 7	& \textcolor{red}{8}	& 9 &	10 & \textcolor{red}{11} & \textcolor{green}{12} \\
\hline
%Keyword used & $OD$ &	$M2$ &	$L3$ & $EV$ &	$L3$ & $M2$ & $EV$ & $EV$ & $L3$ & $EV$ & $OD$ & $M2$ \\
%\hline
%Bits (Alice) & \textsl{1} &	\textsl{0} & \textsl{1} & \textsl{0} & \textsl{1} & \textsl{0} & \textsl{0}	& \textsl{0} & \textsl{1} & \textsl{0} & \textsl{1} & \textsl{0}\\
Bits (Alice) & \textsl{\textcolor{green}{1}} &	\textsl{0} & \textsl{\textcolor{green}{1}} & \textsl{\textcolor{green}{0}} & \textsl{\textcolor{red}{1}} & \textsl{\textcolor{green}{0}} & \textsl{0}	& \textsl{\textcolor{green}{0}} & \textsl{1} & \textsl{0} & \textsl{\textcolor{green}{1}} & \textsl{\textcolor{green}{0}}\\
\hline
\textcolor{blue}{Eve's question} & $nOD$ & $nL3$ & $nL3$ & $nL3$ & $nOD$ & $nOD$ & $nL3$ & $nL3$ & $nOD$ & $nOD$ & $nL3$ & $nOD$ \\
\hline
\textcolor{blue}{Ball's answer} & Yes &	No & Yes & Yes & Yes	& No & No	& Yes & Yes & No & Yes & Yes \\
\hline
\textcolor{blue}{Value of $\mathbf{m}$} & $\mathbf{3}$ &	$\mathbf{3}$ & $\mathbf{1}$ &	$\mathbf{2}$ & $\mathbf{3}$	& $\mathbf{4}$ & $\mathbf{4}$	& $\mathbf{1}$ & $\mathbf{1}$ &	$\mathbf{2}$ & $\mathbf{2}$ & $\mathbf{3}$ \\
\hline
Bob's question & $nOD$ & $nOD$ & $nL3$ & $nOD$ & $nL3$ & $nL3$ & $nL3$ & $nOD$ & $nOD$ & $nL3$ & $nOD$ & $nL3$ \\
\hline
Ball's answer & Yes &	Yes & Yes & No & No	& No & No	& Yes & Yes & Yes & No & No \\
\hline
Bits (Bob) & \textsl{\textcolor{green}{1}} &	\textsl{1} & \textsl{\textcolor{green}{1}} & \textsl{\textcolor{green}{0}} & \textsl{\textcolor{red}{0}} & \textsl{\textcolor{green}{0}} & \textsl{0}	& \textsl{\textcolor{red}{1}} & \textsl{1} & \textsl{1} & \textsl{\textcolor{red}{0}} & \textsl{\textcolor{green}{0}}\\
%Bits (Bob) & \textsl{1} &	\textsl{1} & \textsl{1} & \textsl{0} & \textsl{0} & \textsl{0} & \textsl{0}	& \textsl{1} & \textsl{1} & \textsl{1} & \textsl{0} & \textsl{0}\\
\hline
\end{tabular}}
\hfill{} 
\caption{\textit{Foiling an eavesdropper}. Alice assigns the keywords $EV=M2=\textsl{0}$ and $OD=L3=\textsl{1}$ (first row). Eve intercepts Alice's communication, measures the balls herself and depending on the results, prepares a fresh sequence of balls (with value $\mathbf{m}$) and sends them over to Bob. However, she inevitably makes a mistake, which Alice and Bob later discover for the cases where they used the same basis but don't get the same bit (shown in red). To distinguish between the legitimate users and Eve, the steps performed by the latter are shown in blue (first column).}
\label{tabIRAonMBP}
\end{table*}
\end{center}
Introducing some terminology that shall help as we proceed, we define the process of encoding the state as \emph{preparation}, and the process of questioning a ball as \emph{measurement}. Balls are thus prepared and measured in one of two conjugate pairs (even/odd or more/less) of possibilities; a possibility is technically called a \emph{basis}.
\subsection*{Steps of the protocol}
Below are the steps that demonstrate how Alice and Bob can share a secret key using these magical balls: 
\flushleft
\begin{enumerate}
\item[S1.] Alice encodes `M' (ideally, M approaches infinity) magical balls in the safe environment of her lab or office. The balls are labelled sequentially and the encoding keyword for each ball is carefully recorded. 
\item[S2.] She sends the labelled balls to Bob using normal postal services. Note that the postal services can be \emph{unreliable}, \emph{inefficient} or even \emph{untrustworthy}! 
\item[S3.] Bob receives `W' balls (W$\leq$M, as some could be lost on the way), assembles them sequentially, and \emph{randomly} measures (asks questions a or b to) each ball. He carefully records the corresponding answers. In the end, he obtains a bit sequence comprising of \textsl{0}s and \textsl{1}s (Yes = \textsl{1}, No = \textsl{0}).
\item[S4.] Bob contacts Alice using an authenticated public line\footnote{a communication channel open to passive eavesdropping} and tells her only of the questions he asked, but doesn't divulge the corresponding answers.
\item[S5.] For each ball, Alice compares her encoding keyword (the preparation basis) with the question (the measurement basis) asked by Bob. Whenever there is a match, i.e. \emph{the preparation basis of Alice and the measurement basis of Bob coincide}; Alice can correctly predict Bob's answer. She tells Bob to keep all these (on an average, this would be W/2) instances and discard the rest. In this manner, she finally obtains a bit sequence comprising of \textsl{0}s and \textsl{1}s which is identical to that of Bob. This is the raw secret key! 
\item[S6.] Alice and Bob publicly compare a few randomly-chosen bits from their secret key. If these bits are indeed the same, then they assume the same holds for the rest of the secret key. Discarding the publicly-compared bits, Alice and Bob obtain the final secret key. 
\end{enumerate}
\begin{center}
\begin{table*}[h] %TABLE FOR ANALOGY BETWEEN MBP & BB84
%\centering
{\small
\hfill{}
%\centerline{\includegraphics[width=13cm]{
\begin{tabular}{|c|c|c|}
\hline
 & \textbf{Magical Ball Protocol} & \textbf{Realistic BB84 protocol} \\
\hline
\textbf{carriers of the} & & \\
\textbf{encoded information} & Tiny magical balls & Single photons \\
\hline
\textbf{communication} & & Quantum channel, e.g. optical \\ 
\textbf{medium} & Postal services & fiber or atmospheric link \\
\hline
\textbf{Bases for} & \emph{Mathematical property}: & \emph{Physical property (polarization)}:  \\
\textbf{preparation and} &  Even/Odd ($EV$/$OD$), & Horizontal/Vertical ($\mathbf{H}$/$\mathbf{V}$), \\
\textbf{measurement} &  More/Less ($M2$/$L3$) & Diagonal/Anti-Diagonal ($\mathbf{D}$/$\mathbf{A}$) \\
\hline
\textbf{physical method of} & Balls give answers & Binary system of optical detectors \\
\textbf{realizing bits} & as either Yes or No & that either \emph{click} or don't. \\
\hline
\textbf{Other significant} & Balls can be asked precisely & Single photons get destroyed \\
\textbf{properties} & one question, then fall silent & once they are detected \\
\hline
\end{tabular}}
\hfill{} 
\caption{\textit{Analogy between MBP and realistic QKD protocol}. Alice prepares photons in one of four possible polarizations $\left(\mathbf{H}, \mathbf{V}, \mathbf{D}, \mathbf{A}\right)$ and sends them to Bob on a quantum channel. Bob measures the incoming photons using a system made from optical components and optoelectronic detectors. A successful detection event is registered when an impinging photon causes a click in the detector.}
\label{tabMBPtoBB84}
\end{table*}
\end{center}
The steps described above are illustrated using Table~\ref{tabMBP}. The columns highlighted in green correspond to situations when Alice and Bob used the same basis. If Alice simply assigns $EV = M2 =$ bit \textsl{0} and $OD = L3 =$ bit \textsl{1}, then after discarding instances where dissimilar basis were used, Alice and Bob are left with an identical bitstream. 
\subsection*{Tackling Eve}
To make sure that the magical ball protocol (MBP) works in an adversary scenario as well, Table 3 shows the effects introduced by Eve as she tries to gain knowledge of the key. Eve is assumed to control the postal services, so the balls are essentially in her custody until they are delivered to Bob. Due to assumptions P1--P4, the most suitable methodology for Eve would be to interrogate the balls herself and depending on the answers, encode a new sequence of balls and send them to Bob. 

In roughly half of all the `W' instances, Alice and Bob used the same basis. In about half of these instances, Eve would've used a dissimilar basis. Thus, on average, Eve induces an error in one-fourth or $25\%$ of all the instances, discovering which, Alice and Bob can decide to abort the protocol, discard the key and retry (or perhaps use a different communication link).
\section{Quantum cryptography in practice}
The result obtained in the last line of the previous section is identical to that achieved by the BB84 quantum cryptographic protocol that was proposed by Charles Bennett and Giles Brassard at a conference in Bangalore in 1984~\cite{bb84}. The error rate, called quantum bit error rate (QBER), is the most important quantity relevant to the security of any QKD protocol. To elaborate, at the completion of a QKD protocol, if Alice and Bob incur an error rate above the threshold established by the theoretical security proof (of that protocol), then they discard the key. If the QBER is below threshold but non-zero, then Alice and Bob use privacy-amplification techniques~\cite{pa96} to distill a secret key that reduces Eve's potential knowledge close to zero. An eavesdropper (bounded just by quantum mechanics) is thus destined to be discovered or thwarted and \emph{even an unprecedented amount of computational power cannot help him/her to stay concealed}! 
\begin{figure}[ht]
\centerline{\includegraphics[width=7.5cm]{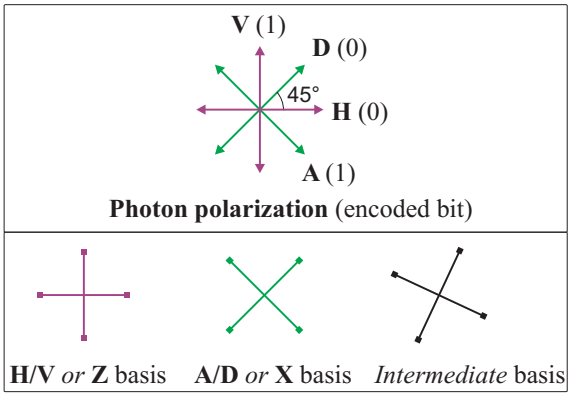}}
\caption{\textit{Bit encoding and polarization bases in typical QKD protocol}.}
\label{figPolnBasis}
\end{figure}

I shall now explain how a typical QKD protocol is implemented in real life. Table 4 strives to make a connection between the realistic implementation of BB84 with the magical ball protocol (MBP) described in the previous section. 

A single photon is indeed the tiny magical ball granted to us by quantum physics. Conjugate bases for preparing and measuring photons obey the \emph{Heisenberg Uncertainty Principle}: only one of the properties from a conjugate pair can be known with certainty. And simultaneously, the knowledge of the other (property) is completely randomized. For QKD, this translates to: given only the outcome of a measurement, one cannot perfectly predict which basis had been applied. This is indeed satisfied by MBP, however, Eve can -- within the framework of quantum mechanics -- resort to an intermediate basis. This carries no classical analogue and thus defines the realm where the functioning of MBP departs from that of a QKD protocol.

In practice, these bases can be easily constructed using the polarization property of photons~\cite{qcrvw02}. Figure~\ref{figPolnBasis} geometrically depicts four polarizations $\mathbf{H}$, $\mathbf{V}$, $\mathbf{D}$ \& $\mathbf{A}$ along with the $\mathbf{H}$/$\mathbf{V}$ basis and the $\mathbf{D}$/$\mathbf{A}$ basis (usually denoted by $\mathbf{X}$ and $\mathbf{Z}$, respectively) that follow the weird characteristics P1--P4 of the magical balls. In a real protocol, Alice starts by encoding classical bits randomly in the $\mathbf{X}$ or $\mathbf{Z}$ basis; physically, it means that photons are randomly encoded in one of four possible polarizations. Note this differs from the MBP, the reasons for that shall become clear later. She then sends the photons over a quantum channel to Bob, who (independently) measures them in the $\mathbf{X}$ or $\mathbf{Z}$ basis.

\begin{figure*}
\centerline{\includegraphics[width=13.4cm]{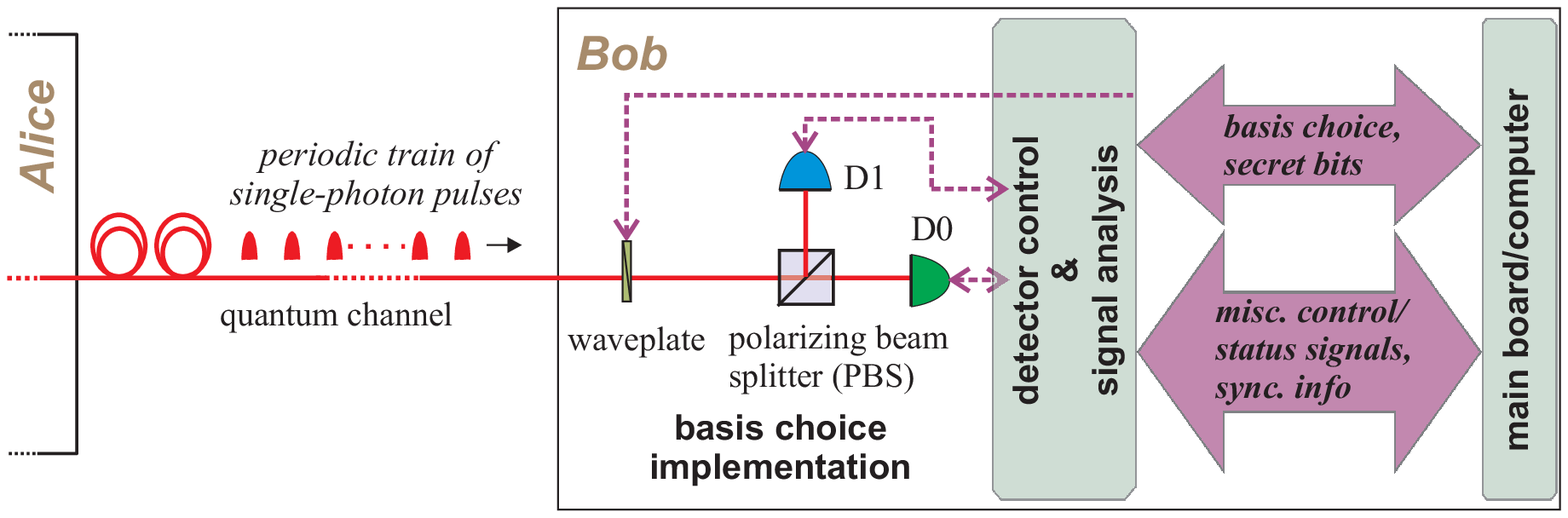}}
\caption{\textit{Schematic of a typical measurement setup for polarization based QKD}. Bob randomly chooses $\theta = 0^{\circ}$ or $22.5^{\circ}$ to apply the $\mathbf{Z}$ or $\mathbf{X}$ basis, respectively. Depending on the incoming state of polarization, a single photon gives a click in either of the two detectors. E.g. to communicate bit \textsl{0}, Alice sends either a horizontally- or diagonally- polarized ($\mathbf{H}$ or $\mathbf{D}$) photon. If the waveplate is set at $\theta = 0^{\circ}$, the $\mathbf{H}$ photon simply transmits through the PBS and gives a click in D0. But in this situation, a $\mathbf{D}$ photon would be transmitted or reflected by the PBS with an equal probability, so it could give a click in either D0 or D1. However, if the waveplate is set at $\theta = 22.5^{\circ}$, then the $\mathbf{D}$ photon is rotated and becomes an $\mathbf{H}$ photon, thus getting transmitted through the PBS and yielding a click in D0.}
\label{figBobdetection}
\end{figure*}

The measurement outcomes are obtained using a specialized optical detection assembly made from single-photon detectors (SPDs). One such possible detector assembly in Bob's device is shown in Fig.~\ref{figBobdetection}. The basis choice is implemented by a waveplate and polarizing beam splitter (PBS). A waveplate rotates the polarization of the input photon depending on the relative angle $\theta$ between its optic axis and the input polarization. A PBS on the other hand allows a single photon to be transmitted [reflected] if the photon is in a horizontal [vertical] polarization. The photon is then detected by D0 [D1], where the latter numeral signifies the measurement outcome -- the classical bit obtained. \linebreak[1]
Some key features that apply to a majority of the currently-known QKD systems are summarized below: 

R1. \textbf{Periodic operation}: The sender (usually Alice) emits optical signals, such as pulses of single photons, at well-defined times. \\
R2. \textbf{Basis choice application}: In theory, basis choice (either preparation or measurement) happens at a single instant of time. In practice however, this is usually implemented by application of an electronic voltage pulse which (obviously) has a finite rise and fall time. \\
R3. \textbf{Dead time of the single-photon detectors}: After a detection event, an SPD is rendered inactive for a short period of time. This is known as dead time and normally ranges from a few nanoseconds (ns) to a few microseconds ($\mu$s). \\
R4. \textbf{Finite efficiency of detection and dark counts}: Not every single-photon pulse that impinges on a SPD produces a click. Owing to technological limitations, typical single-photon detectors employed in QKD systems click only $10\!-\!15\%$ of the times; this no$.$ is described as the \emph{detection efficiency} and is usually denoted by the Greek letter $\eta$. Moreover, a detector may click even in complete absence of optical input; this is known as a \emph{dark count}. \linebreak[2]

The probability of obtaining a detection click from dark counts is typically smaller by a factor of $10^5$ or more than from a single photon. Nonetheless, a legitimate bit $\textsl{0}$ then requires not just a click in D0, but also no-click in D1 (and vice-versa for bit $\textsl{1}$).  
\begin{figure}[h]
\centerline{\includegraphics[width=7.5cm]{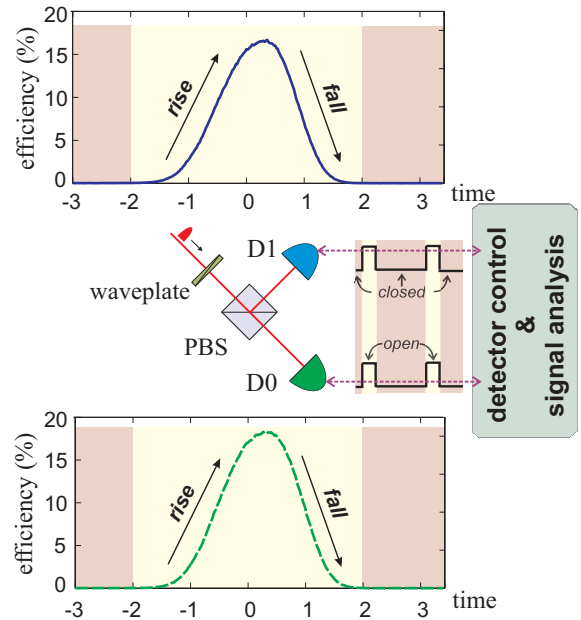}}
\caption{\textit{Detection efficiency as a function of time}. The detector control \& signal analysis circuit sends a voltage pulse to each of the detectors. Due to the finite response times, the corresponding efficiency curves rise and fall in a shorter interval. The peak value achieved depends on the optical \& electrical properties of the detector material, the applied electronic voltage etc.}
\label{figNormeff}
\end{figure}

Since SPDs operate at extremely low levels of light, they are activated only for the short duration when the periodic single-photon pulses from Alice are expected to arrive. The detector control, utilizing synchronization info (refer Fig. \ref{figBobdetection}), electronically opens a window on both detectors at the time a single photon \emph{may} impinge upon them. For the remaining time, the window is kept closed and the detectors are not much sensitive to stray light. The information about the measurement outcome in this time-window, i.e. ``which detector clicked?'' is subsequently passed onto the main board/computer. 

In Fig.~\ref{figNormeff}, a typical `windowed mode' optical detection that considers R1--R4 is depicted.
\section{Quantum hacks \& attacks based on timing}
To obtain a tangible security primitive, e.g. the secret key bitstream, any physical QKD system thus needs to make a \emph{transition from the quantum to classical domain}. Quantum physics may prove that a QKD protocol, operated under certain assumptions, is secure. But can practical implementations also certify that beyond doubt? 
\begin{figure}
\centerline{\includegraphics[width=7.5cm]{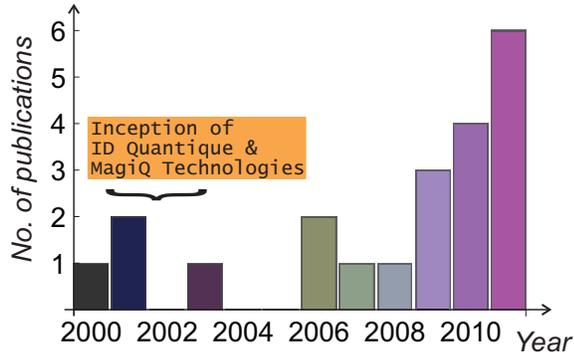}}
\caption{\textit{Rising popularity of quantum hacking}. The bar chart shows the growing no. of publications devoted to (experimentally) finding and exploiting limitations of practical QKD systems. Only papers from peer-reviewed journals have been considered, so this is by no means an exhaustive list. ID Quantique (http://www.idquantique.com) and MagiQ Technologies (http://www.magiqtech.com) were the first two companies to commercialize QKD.}
\label{figpubsQHckg}
\end{figure}
A quest for the answer to this question began roughly a decade ago and has led to some astonishing results~\cite{leuchs11}; see Fig.~\ref{figpubsQHckg}. Termed `quantum hacking', this research field has witnessed many successful proof-of-principle attacks devised and performed on practical QKD systems. The attacks primarily show how an eavesdropper obtains partial or full info about the secret key without breaching the QBER threshold. It should be stressed that a majority of the eavesdropping strategies utilized differences between the security proof of the QKD protocol (a.k.a. the theoretical model) and the actual implementation. These differences mainly arise due to technical imperfections or deficiencies of the hardware, such as single-photon detectors.

The main topic of this article is quantum hacking activities that fall under the gamut of timing attacks. In the following, I shall discuss a few (published) works that share the common theme of exploiting time as a parameter or variable to compromise the security of QKD. I shall, while resorting to the minimum possible technical details, elaborate the following: 
\flushleft
\begin{itemize}
\item Main exploitable vulnerability
\item	Principle of the attack
\item	Result and impact
\end{itemize}
General \& specific countermeasures to prevent/detect these attacks and implications of quantum hacking on QKD will be discussed in the end.

\subsection{Effects of detector efficiency mismatch on security of quantum cryptosystems}\label{qh:dem}
\begin{figure}
\centerline{\includegraphics[width=7.6cm]{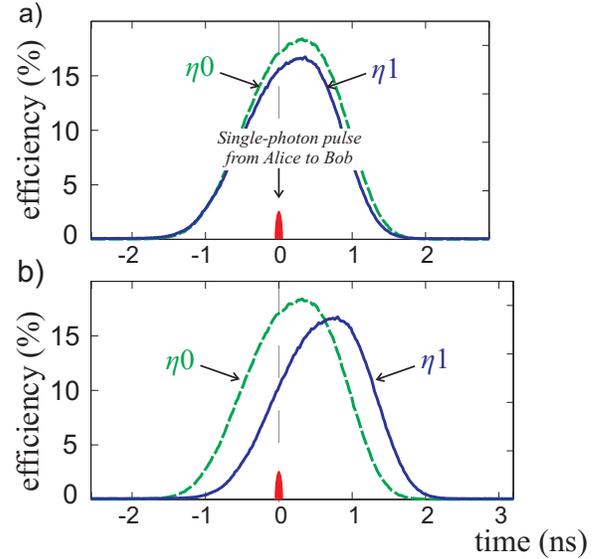}}
\caption{\textit{Temporal detector efficiency mismatch}. a) Apart from a minor difference in the peak value and overall shape, the efficiencies overlap each other quite well, or temporal mismatch is negligible. This would be alright for a practical QKD system. b) The mismatch is rather pronounced and could arise due to electrical differences, e.g. the length of the circuit wire to D0 may be slightly shorter than the one to D1, so the ``window open'' pulse dispatched by the detector control reaches D0 roughly 0.4 ns before D1. As a result, the single-photon pulse in b) sees a vastly different efficiency of detection.}
\label{figDEM}
\end{figure}
For the correct working of a QKD protocol the measurement outcome should be random [deterministic] whenever the preparation basis is different than [equal to] the measurement basis. It entails here that to make sure this holds true for practical QKD, it is mandatory that the chances of detection by D0 and D1 are equally likely, i.e. any discrimination between the detectors should not be possible. However, realistically speaking, it is impossible to construct two identical detectors. Even if one manages to make them very similar, their respective detection windows -- while individually being open during the arrival time of a photon -- are still prone to shift relative to each other. This could happen, e.g. due to finite manufacturing tolerances, such as tiny optical path length differences or circuit length differences, which fluctuate because of temperature variations etc. and thus are very hard to control. In such a situation, the system exhibits a \emph{detection efficiency mismatch} in time and this was the main topic of a paper~\cite{dem06} by researchers from the Norwegian University of Science and Technology (NTNU) and St. Petersburg State Polytechnic University. Figure~\ref{figDEM} illustrates the case of two detectors in Bob responding differently to single photons that impinge upon them at the same instant of time. 

Furthermore, the researchers also theoretically proposed a special kind of \emph{intercept-and-resend} strategy (see Table~\ref{tabIRAonMBP}) to eavesdrop in such a scenario. Eve intercepts and measures the states sent by Alice on the quantum channel to Bob. Depending upon a measurement outcome, she resends a new state prepared in such a manner that it would produce a detection event in Bob, if and only if his measurement basis is opposite to Eve's preparation basis. In this way, Eve imposes faked detection events in Bob and hence, such an attack is referred to as the \emph{faked-states attack}. 

Although this paper was primarily a theoretical study, i.e. it did not demonstrate an attack on a QKD system as a whole, yet it is considered as a landmark paper because its main ideas have formed the core of countless eavesdropping strategies in last 5 years. In fact, two of the attacks that I shall discuss further on are based on detector efficiency mismatch (DEM) and two utilize the concept of faked states! 

\subsection{Breaking a quantum key distribution system through a timing side channel}\label{qh:tsc}
It should be clear how critical it is for the detectors of a QKD system to appear indistinguishable to the outside world (especially Eve). Researchers from National University of Singapore in 2007 showed how timing information revealed during public discussion between Alice and Bob (S4 \& S5 in the protocol, section~\ref{sec:mbp}) could allow an adversary to discern between the detectors and hence, eavesdrop without introducing perceptible errors. 

\begin{figure}
\centerline{\includegraphics[width=7.5cm]{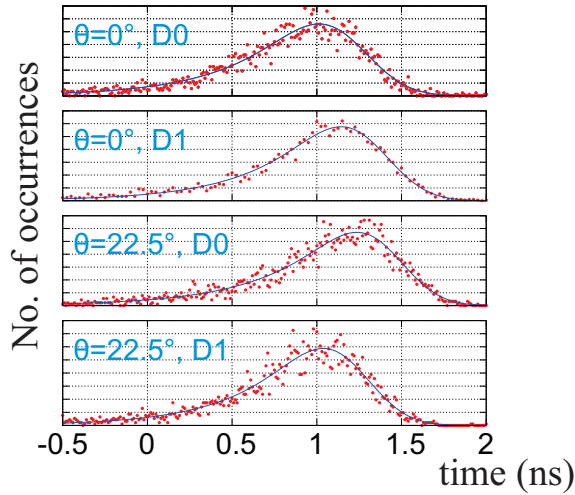}}
\caption{\textit{Timing side channel attack}. Timing histograms collected from all four measurement possibilities (blue labels) show different centroid locations. At angle $\theta = 22.5^{\circ}$ ($\mathbf{X}$ basis) the separation between peaks of D0 and D1 is especially prominent; once the basis choice is revealed on the public channel, Eve has some chance to know if the obtained bit was \textsl{0} or \textsl{1}. Figure reproduced with permission, \copyright2007 Optical Society of America.}
\label{figTSC}
\end{figure}

As such it seems implausible to think of a correlation between measurement outcomes and publicly exchanged timing information (e.g. synchronization timestamps). However, if the timing info (between Alice and Bob) is communicated with a high resolution, an eavesdropper could perhaps know the answer to \emph{which detector clicked} just by carefully scrutinizing the timestamps! Figure~\ref{figTSC} depicts a histogram of the optical detection timings from the compromised QKD system~\cite{tsc07} and conveys the idea of this attack. 

Technically, this may be due to the same reasons that cause the detector efficiency mismatch (explained in the previous attack). More qualitatively, this should be understood as \emph{inadvertent} encoding of information about the secret key in undesired degrees of freedom. These degrees of freedom are formally called \emph{side channels} and have been extensively explored in classical cryptography. 
The info about the secret key that Eve could obtain with such an attack was also evaluated theoretically. Detection systems uncompensated by a mere 0.5 ns -- an amount that can go easily unnoticed in usual experimental setups -- could allow the eavesdropper an access $\geq$ 25\% of the key! Thus, the loophole exposes an important issue which cannot be overlooked.

\subsection{Experimental demonstration of time-shift attack against practical quantum-key-distribution systems}\label{qh:tsa}
\begin{figure}
\centerline{\includegraphics[width=7.7cm]{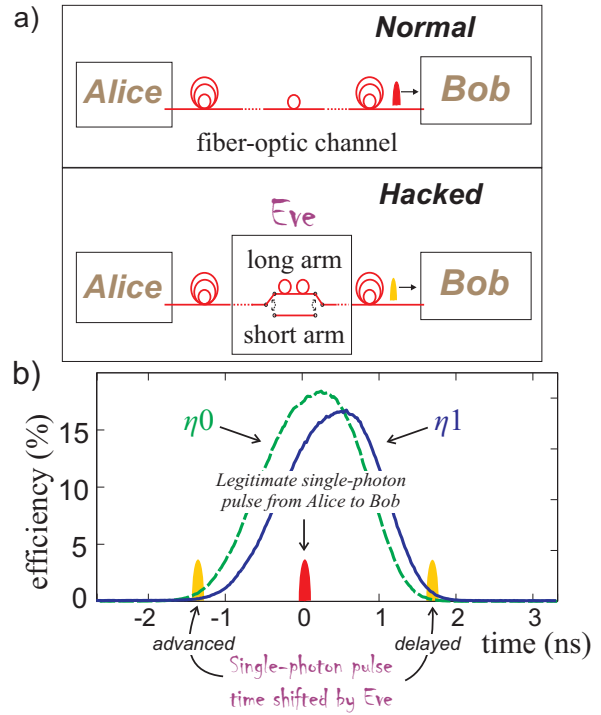}}
\caption{\textit{Time-shift attack}. To eavesdrop, Eve simply shifts -- advances or delays -- the legitimate optical signals from Alice that are on the way to Bob. This could be achieved by using an assembly of optical switches and delay fibers as shown in the Hacked case.}
\label{figTSA}
\end{figure}

Researchers from University of Toronto made an attempt to exploit the aforementioned DEM vulnerability by performing a time-shift attack~\cite{tsa08}. This was the first known (proof-of-principle demonstration of an) attack on commercial QKD, the system in this case being manufactured by the Swiss company ID Quantique. The idea behind the attack, illustrated in Fig.~\ref{figTSA}, is quite simple. Eve cuts a section of the quantum channel and places an assembly of optical switches that reconnect Alice and Bob via two fibers, one \emph{shorter} and other \emph{longer} than the cut section. 

From the viewpoint of \emph{actual} execution, the time-shift attack (TSA) unfortunately suffered from numerous problems: \\
1. To apply the attack, the temporal mismatch needs to stay high, i.e. $\eta0$ and $\eta1$ peaks should be far apart, throughout the running of the protocol. As reported by the authors of this work, this occurred quite rarely (4\% of all the instances).\\
2. The mismatch not only occurs infrequently, but is also probabilistic. The efficiency $\eta0$ could be left, right or on top of $\eta1$ but Eve \emph{cannot predict} this at all. The QKD system also does not reveal any information that could facilitate Eve in this regard. \\
3. The magnitude of the highest mismatch is also not exceptional. Consequently, Eve has to displace the states to the extremities of the window for executing TSA. The overall detection rate seen by Bob (and Alice) would then be much lower than expected, this could raise alarm.

Thus, the efficacy of eavesdropping is considerably reduced. Nonetheless, the TSA was instrumental in bringing realistic QKD into limelight again and spurring a slew of quantum hacking activities. 

\subsection{After-gate attack on quantum cryptosystem}\label{qh:afg}
\begin{figure}
\centerline{\includegraphics[width=7.8cm]{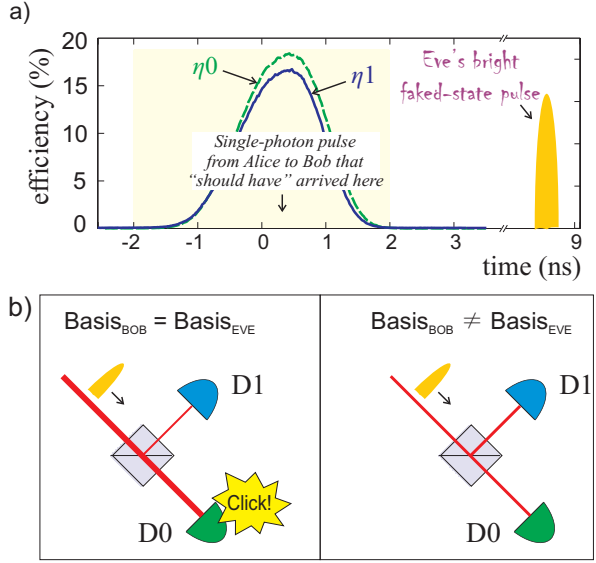}}
\caption{\textit{After-gate attack}. Eve launches bright pulses that arrive outside Bob's detection window, yet they make his detectors click according to her will.}
\label{figAFG}
\end{figure}
In 2010, researchers from NTNU, Max Planck Institute for the Science of Light (MPL) and Universidad de Guanajuato had experimentally demonstrated the blinding of single-photon detectors (SPDs) in two different commercial QKD systems~\cite{lars10}. This has perhaps been the most powerful and the best-performing hack on QKD so far. In 2011, they targeted yet another imperfection of these SPDs and based on the idea of faked states, they were able to remotely control the measurement outcome in Bob~\cite{carlos11}. 
The limitation that they exposed was that an SPD can be \emph{forced to click} when a sufficiently bright pulse (typically containing millions of photons) hits it in the dead time, i.e$.$ outside the detection window (see Fig.~\ref{figAFG}a). The main principle of the attack is as follows: 

1. Eve intercepts the legitimate quantum states from Alice and measures them. This could be done using an exact copy of the Bob device. Let us assume she uses basis $\mathbf{X}$ and detects bit \textsl{0}. \\
2. She now prepares a bright pulse in the $\mathbf{X}$ basis and times its arrival in Bob such that:
\begin{itemize}
\item	If Bob chooses the same basis as Eve, all optical power from the bright pulse is diverted to D0. 
\item	If Bob chooses the other basis, then the power splits equally among both the detectors. 
\end{itemize}
It turns out that for the SPDs in Bob, there is an optical power threshold for bright pulses above [below] which they would always [never] click! This situation is illustrated in Fig.~\ref{figAFG}b.

Due to impinging of bright pulses on a SPD, a phenomenon called \emph{afterpulsing} is encountered in practice. This increases dark counts (and hence the QBER) which could deter Eve, but using some clever strategies, the researchers were still able to simulate an attack that showed the possibility of eavesdropping with the total $\rm{QBER} < 11\%$ at the operating conditions of the commercial system.

\subsection{Device calibration impacts security of quantum key distribution}\label{qh:llm}
\begin{figure}
\centerline{\includegraphics[width=7.8cm]{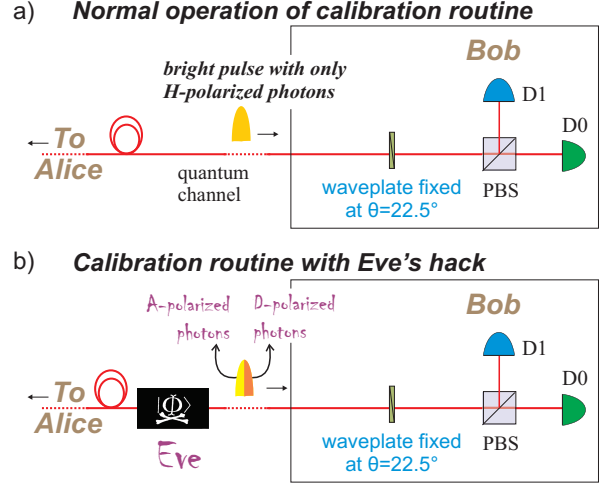}}
\caption{\textit{Hacking QKD calibration to induce DEM}. Eve spoofs the calibration pulses such that they still produce the expected no. of clicks in D0 and D1 while the detection windows are being scanned, but with a relative delay which induces a temporal DEM. }
\label{figLLM}
\end{figure}
It has been assumed that Alice sends single-photon pulses and Bob opens a narrow detection window \emph{at the appropriate time} to measure them. But how is such synchronization obtained in the first place? QKD systems usually run a \emph{calibration routine} for this purpose: Alice and Bob exchange timing information to calibrate their devices, before any secret key exchange. The commercial QKD system from ID Quantique implements it (conceptually) as follows: \\
1. Alice prepares relatively-bright pulses (containing a few $1000$s of photons) in the $\mathbf{H}$ polarization (bit \textsl{0} encoded in $\mathbf{Z}$ basis) and sends them to Bob.\\
2. Bob \emph{always} measures in the $\mathbf{X}$ basis. Every photon in the pulse thus has an equal chance to yield a click in either D0 or D1.\\
3. Bob temporally scans the detection windows for D0 and D1 over the incoming pulses while monitoring the no. of clicks at each scan position.

By independently maximising the no. of clicks in each detector, Bob not only knows becomes synchronized to Alice, but also compensates for any detector efficiency mismatch (DEM). Figure 10a illustrates the optical scheme. 
Researchers, primarily from MPL and NTNU, found a weakness in this calibration routine and experimentally demonstrated a way to hack it~\cite{Jain11}; see Fig.~\ref{figLLM}b. Before the calibration pulses reach Bob, Eve manipulates them such that the photons in the first temporal half of each pulse are $\mathbf{D}\!-\,$polarized (yielding clicks in D0) and in the latter half are $\mathbf{A}\!-\,$polarized (yielding clicks in D1). An artificial displacement of the efficiencies is thereby \emph{induced}, resulting in both large \& deterministic DEM as already depicted in Fig.~\ref{figDEM}. 

\subsection{Other attacks \& countermeasures}\label{qh:oac}
The attacks described in above are of course not the only ones proposed/performed on practical QKD systems. Due to space constraints, I picked what all could be most easily tagged with \emph{timing attacks}. The interested reader is invited to explore contextually similar works, e.g. the phase remapping attack~\cite{Xu10} and the dead time exploitation of single-photon detectors~\cite{Weier11}. 

The detector efficiency mismatch (DEM) problem outlined in $\ref{qh:dem}$ has clearly enabled several timing-based attacks on QKD so far. It allowed eavesdropping via time-shift attack ($\ref{qh:tsa}$), faked-state attack ($\ref{qh:llm}$), or through exposed timing side channels ($\ref{qh:tsc}$). By carefully equalizing delays, randomizing basis choice and truncating precisions, DEM and associated vulnerabilities can be minimized and these attacks could be defeated. However, in general, a QKD system should still characterize timing-related information and subsequently account for any potential leakage of the secret key to Eve. Installation of watchdog detectors and optical isolators in Bob could thwart off bright pulse attacks ($\ref{qh:afg}$ and $\ref{qh:llm}$), however, they also decrease the legitimate throughput. Even more, such technical plugs \& countermeasures that appear intuitively correct might (however) give rise to new and unthought-of loopholes. Hence, it is vital to carefully adapt them to the quantum protocol. 

Another recent approach that advocates an altogether different paradigm of performing QKD is \emph{device independence}: no knowledge of how the physical devices (of Alice and Bob) function is required! The only necessities are that quantum physics is correct and that Alice and Bob fully control the signals emitted and received by their devices~\cite{diqkd07}. Some new \& exciting proposals in this direction have already claimed to get rid of most known detector flaws and side channels~\cite{mdiqkd12a, mdiqkd12b}.

\section{Conclusion}
In this article, I've discussed a coterie of attacks against QKD systems. There is, however, a vast (and growing) literature -- explored from both fundamental and technical viewpoints -- on security concerns in practical QKD. The good news is that so far none of the attacks have been proven insurmountable. So, if we believe in laws of quantum physics, then we can also believe that QKD has a future. It is already speculated that public-key cryptography would be in great danger if and when quantum computers become a reality. In light of that, QKD is perhaps the best and most viable alternative. Nonetheless, it must be stressed that the role played by physics guarantees \emph{merely} a single link in the whole chain of security. In an excellent paper, very appropriately titled the black paper of quantum cryptography~\cite{blackpaper09}, the authors write:
\begin{quote}
\emph{We believe that this state of affairs cannot be simply dismissed with a ``there have been examples of bad design of the device''. At any stage of development, the devices were actually carefully designed; the security claims of the authors were accepted as valid by referees and colleagues. Neither now, nor at any later time, will one be able to guarantee that the devices in use are provably good. And it is certainly not a good idea to close one's eyes, invoke the laws of physics and dump on them a responsibility they cannot possibly bear.}
\end{quote}
At the beginning of section 4, the question ``Will realistic systems that execute a quantum cryptographic protocol be able to certify its security claim beyond doubt?'' was posed. The above quote provides a very suitable answer to that. So, a reasonable and acceptable level of security is definitely achievable, but like any other technology, practical QKD must also undergo a tight scrutiny to weed out vulnerabilities and patch loopholes. 

Finally, I would like to remark that there are innumerable aspects of practical QKD that could not be covered by this article. These range from efforts on standardization of QKD~\cite{qkdstdzn} to integrating it into public networks~\cite{uqcc10} and even taking it into outer space using satellites~\cite{qkdsat}. All in all, the take-home message is that quantum cryptography is quite a vibrant field with a tremendous potential for providing a secure communication technology in the future.

\end{document}